\documentclass[reprint,
superscriptaddress,
 amsmath,amssymb,
 aps,
pra, letterpaper,
square,comma,numbers,sort&compress,doi=false,url=false
]{revtex4-1}
\usepackage{natbib}
\usepackage{graphicx}
\usepackage{amsmath}
\usepackage{upgreek}
\usepackage{hyperref}
\usepackage{gensymb}
\usepackage{tabularx} 

\newcommand{\um}{$\upmu\mathrm{m}\;$}

\newcommand{\umdot}{$\upmu\mathrm{m}$}

\usepackage{epstopdf}
\usepackage[version=3]{mhchem}

\begin{document}

\title{Optically coherent nitrogen-vacancy centers in \umdot-thin etched diamond membranes}

\author{Maximilian Ruf}
\thanks{These authors contributed equally to this work.}
\affiliation{QuTech, Delft University of Technology, 2628 CJ Delft, The Netherlands}
\affiliation{Kavli Institute of Nanoscience, Delft University of Technology, 2628 CJ Delft, The Netherlands}

\author{Mark IJspeert}
\thanks{These authors contributed equally to this work.}
\affiliation{QuTech, Delft University of Technology, 2628 CJ Delft, The Netherlands}
\affiliation{Kavli Institute of Nanoscience, Delft University of Technology, 2628 CJ Delft, The 
Netherlands}
\affiliation{Current address: Clarendon Laboratory, Parks Road, Oxford OX1 3PU, United Kingdom}

\author{Suzanne van Dam}
\affiliation{QuTech, Delft University of Technology, 2628 CJ Delft, The Netherlands}
\affiliation{Kavli Institute of Nanoscience, Delft University of Technology, 2628 CJ Delft, The Netherlands}

\author{\mbox{Nick de Jong}}
\affiliation{QuTech, Delft University of Technology, 2628 CJ Delft, The Netherlands}
\affiliation{Netherlands Organisation for Applied Scientific Research (TNO), 2628 CK Delft, The Netherlands}

\author{Hans van den Berg}
\affiliation{QuTech, Delft University of Technology, 2628 CJ Delft, The Netherlands}
\affiliation{Netherlands Organisation for Applied Scientific Research (TNO), 2628 CK Delft, The Netherlands}

\author{Guus Evers}
\affiliation{QuTech, Delft University of Technology, 2628 CJ Delft, The Netherlands}
\affiliation{Kavli Institute of Nanoscience, Delft University of Technology, 2628 CJ Delft, The Netherlands}

\author{Ronald Hanson}
\email{r.hanson@tudelft.nl}
\affiliation{QuTech, Delft University of Technology, 2628 CJ Delft, The Netherlands}
\affiliation{Kavli Institute of Nanoscience, Delft University of Technology, 2628 CJ Delft, The Netherlands}

\begin{abstract}
Diamond membrane devices containing optically coherent nitrogen-vacancy (NV) centers are key to enable novel cryogenic experiments such as optical ground-state cooling of hybrid spin-mechanical systems and efficient entanglement distribution in quantum networks. Here, we report on the fabrication of a (3.4 $\pm$ 0.2) \um thin, smooth (surface roughness r$_q$ $<$ 0.4 nm over an area of \mbox{20 \um} by 30 \umdot) diamond membrane containing individually resolvable, narrow linewidth ($<$ 100 MHz) NV centers. We fabricate this sample via a combination of high energy electron irradiation, high temperature annealing, and an optimized etching sequence found via a systematic study of the diamond surface evolution on the microscopic level in different etch chemistries. While our particular device dimensions are optimized for cavity-enhanced entanglement generation between distant NV centers in open, tuneable micro-cavities, our results have implications for a broad range of quantum experiments that require the combination of narrow optical transitions and \umdot-scale device geometry.

\end{abstract}

\maketitle

The negative nitrogen-vacancy (NV) center is a point defect center in diamond~\cite{Doherty2013,Atature2018} that is used in a wide range of experiments, including quantum sensing~\cite{Degen2008,Maze2008,Balasubramanian2008,Grinolds2011,Maletinsky2012}, quantum computation algorithms~\cite{Waldherr2014, Cramer2016}, and quantum communication~\cite{Hensen2015,Kalb2017,Humphreys2018}. In addition to second-long spin coherence times~\cite{Abobeih2018} and spin-conserving optical transitions~\cite{Robledo2011}, NV centers feature coupling to nearby nuclear spins that can act as memory quantum bits~\cite{Maurer2012,Taminiau2014,Kalb2017}. Many NV-based experiments require a combination of good optical and spin properties in nano-fabricated structures; these experiments include Purcell enhancement of the optical zero-phonon line (ZPL) transitions in a diamond micro-cavity~\cite{Englund2010,Wolters2010,VanDerSar2011,Faraon2012,Hausmann2013,Lee2014,Li2015,Riedrich-Moller2015,Faraon2011,Barclay2011,Gould2016,Kaupp2013,Johnson2015,Riedel2017a} for entanglement generation speed-up, optical ground state cooling of a hybrid NV-cantilever spin-mechanical system~\cite{Rabl2010,Arcizet2011,Kolkowitz2012,Kepesidis2013,Golter2016,Lee2016}, and resonant optical readout of NV centers in sensing applications~\cite{Robledo2011}. While good spin coherence has been demonstrated for surface-proximal NVs (depth of $\sim$ 50 nm)~\cite{Ohno2012,Kim2012,McLellan2016}, the incorporation of optically coherent NV centers in \umdot-scale devices remains an outstanding challenge.

The optically excited state of the NV center is sensitive to electric fields and crystal strain~\cite{Doherty2011}. Therefore, high-frequency electric field noise can lead to dephasing in the excited state, while on longer timescales the transitions can be effectively widened by slow spectral diffusion originating from a changing charge state distribution in the environment~\cite{Fu2009,Siyushev2013}. Although the latter effect can be mitigated by actively tracking the transition frequencies and adding feedback~\cite{Hensen2015}, it comes at the cost of reduced experimental repetition rates. Dephasing poses a more fundamental challenge: in hybrid-mechanical systems, effective optical ground state cooling requires operation in the sideband resolved regime~\cite{Rabl2010,Kepesidis2013}. Dephasing also determines the resolvable magnetic field changes in sensing experiments, and limits the two-photon quantum interference contrast that translates into state fidelity for entanglement protocols~\cite{Bernien2012,Bernien2013}. For all discussed applications of NV centers in membranes, spectral diffusion widths $<250$ MHz and dephasing widths $< 100$ MHz are in practice desired. However, reported spectral diffusion widths in thin ($\sim$ 1 \umdot) nanofabricated structures are $\sim$ 1 GHz~\cite{Riedel2017a} for NV centers formed via nitrogen implantation~\cite{Pezzagna2011}. Improvements in fabrication and preparation methods are therefore necessary to produce devices with linewidths sufficiently narrow for the experiments discussed above.

For our desired application of embedding a diamond membrane in an open, tuneable Fabry-Perot microcavity to increase entanglement generation rates between distant NV centers~\cite{Kaupp2013,Johnson2015,Janitz2015,Riedel2017a,Bogdanovic2017}, we target a final membrane thickness of $\sim$ 4 \umdot. This choice is a compromise between low cavity mode volume needed for high Purcell enhancement and sufficient thickness to avoid frequent breaking of membranes during sample handling, whilst being able to embed NVs as deep in the diamond lattice as possible to avoid surface-induced noise. We furthermore require smooth samples (r$_q <$ 0.3 nm over the $\sim$ 4 \umdot$^2$ large area of the cavity beam waist) to limit losses due to scattering at the diamond-air interface~\cite{Janitz2015,VanDam2018}.

\begin{figure}
	\includegraphics[scale=0.35]{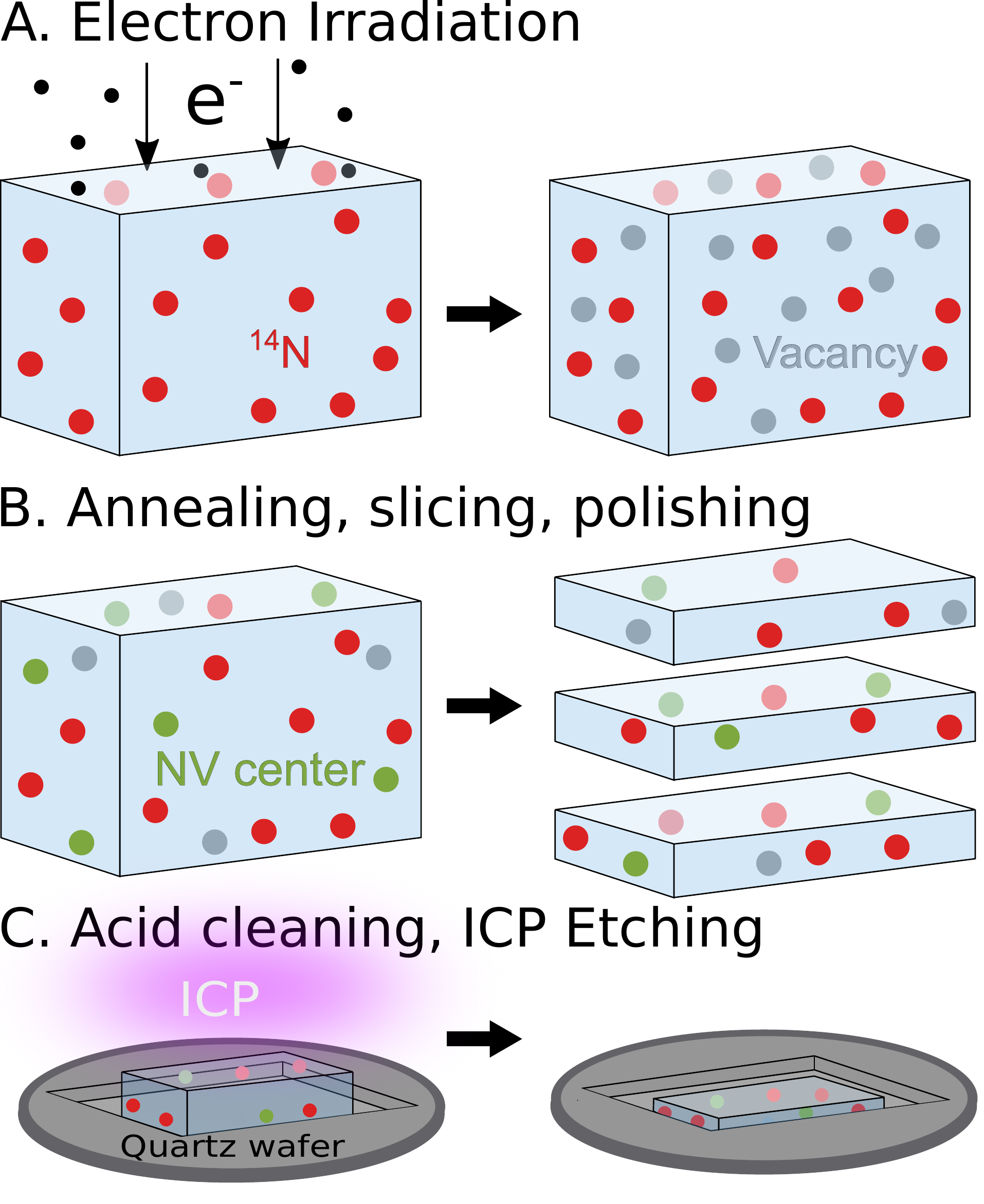}
	\caption{Schematic of the diamond fabrication process flow. (A) A CVD grown diamond is irradiated with electrons to form vacancies in the diamond lattice~\cite{supplement}.
	(B) After a tri-acid clean, a high temperature annealing process in high vacuum combines the holes with naturally occurring nitrogen in the sample to form NV centers~\cite{supplement}. The sample is then sliced into three 50 \um thick slabs that are each polished to a surface roughness of typically r$_q$ $<$ 1 nm. 
	(C) After an acid clean to remove remaining contamination that could lead to masking during etching, the sample is etched in a ICP-RIE.}
	\label{fgr:fabflowandsetup}
\end{figure}

Here, we report on a full fabrication procedure that combines high energy electron irradiation, high temperature annealing, and an optimized etching sequence, to yield a diamond device that meets all the above requirements. We verify the desired NV center optical properties using an extensive study on linewidths following different etching steps.

We start from ultrapure, CVD grown diamond that contains only a few NV centers~\cite{supplement}, and thus requires their density to be increased. NV centers created from implanted nitrogen atoms were recently found to predominately feature broad optical lines, likely due to associated diamond lattice damage~\cite{vanDam2018_implanted_coherence}. Instead, we use high energy electron irradiation to create vacancies throughout the whole diamond that can form NV centers with native nitrogen~\cite{Campbell2000} (see Fig.~\ref{fgr:fabflowandsetup} (A)). After tri-acid cleaning, a high temperature and high vacuum annealing sequence --- consisting of three temperature steps --- ensures the recombination of vacancies with naturally occurring nitrogen in the diamond to form NV centers, and anneals out vacancy chains~\cite{Orwa2011,Chu2014} (see Fig.~\ref{fgr:fabflowandsetup} (B)). The resulting density of NV centers thus depends on the nitrogen distribution resulting from diamond growth, and the number of vacancies created during electron irradiation~\cite{supplement}. The diamonds are subsequently sliced and polished into three membranes of $\sim$ 50 \um thickness each. This value is a trade-off between ease of handling of the membranes in further processing steps, and the amount of material that needs to be removed in a subsequent reactive ion etching (RIE) step using an Inductively Coupled Plasma (ICP) (see Fig.~\ref{fgr:fabflowandsetup} (C)).

To find an etch recipe that leads to smooth, thin diamonds, we investigate the effect of different etch chemistries on the diamond surface roughness on the microscopic level. It is advantageous to employ an oxygen-based ICP-RIE, due to reported high diamond etch rates ($>$ 200 nm$/$min~\cite{Hausmann2010}), and an oxygen-rich surface termination of the diamond, which is beneficial for the charge state stability of the NV$^-$ center~\cite{Petrakova2012}. However, it has been speculated that particles, e.g.~those introduced during the diamond polishing process, can lead to hole formation during etching with O$_2$ \cite{Grillo1997,Atikian2014}. 

\begin{figure}
	\includegraphics[scale=0.38]{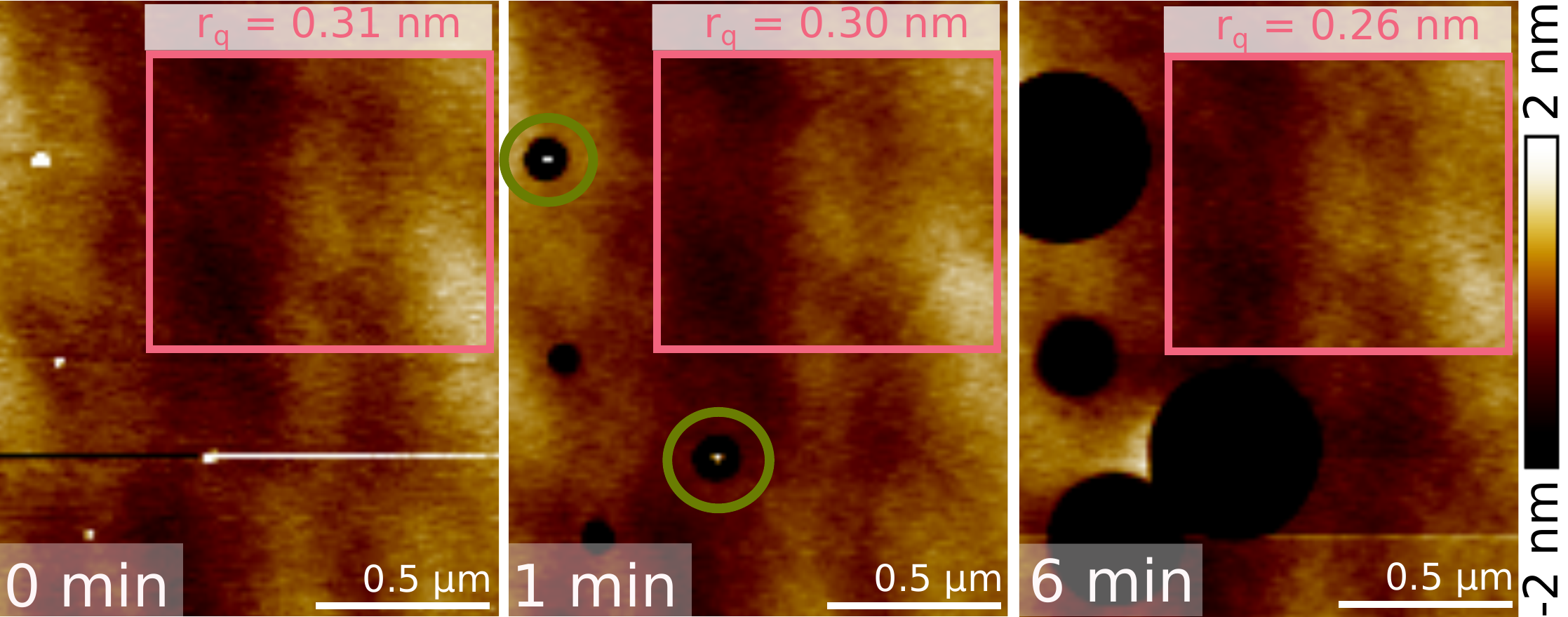}
	\caption{Evolution of a diamond surface during ICP-RIE with O$_2$ for different etching times (indicated in the bottom left corner of each AFM image for an identical area). The data demonstrates that debris on the diamond surface leads to micro-masking during etching, resulting in the formation of nanopillars (visible in the green encircled parts after 1 minute of etching). These pillars are subsequently removed via an isotropic etch component. Pink rectangles show a comparison of surface roughness r$_q$ for a 1 \um by 1 \um wide area that excludes etch-induced holes (error 0.03 nm). 
	Note that the black / white trace in the AFM image before etching is a data acquisition artifact.}
	\label{fgr:o2evolution} 
\end{figure}

By overlaying Atomic Force Microscope (AFM) surface images of identical diamond areas before and after O$_2$ etching, we find that each of the observed circular pits originates from a particle that was initially present at that location. Fig.~\ref{fgr:o2evolution} shows the evolution of one such area before etching, and after 1 and 6 minutes of etching with O$_2$~\cite{supplement}. The data demonstrates that the underlying mechanism of hole formation is micro-masking: the particle etch rate is lower than that of the bulk diamond surface. This leads to the formation of diamond nanopillars on the surface that deflect the impinging plasma, which enhances the etch rate locally and thus creates a hole around the pillar~\cite{Hausmann2010} (see the green encircled areas in Fig.~\ref{fgr:o2evolution} after 1 minute of etching). Due to an isotropic etch component, the pillars are eventually etched away, leaving behind a hole. These holes then remain, and widen as the etching continues. Importantly, we also find that the membrane surface roughness can be maintained during O$_2$ etching if particle-induced holes are excluded (see the pink rectangles in Fig.~\ref{fgr:o2evolution}). Even after extended O$_2$ etches, one thus expects to maintain the intitial diamond surface roughness if particles can be effectively removed before this step.

\begin{figure}
	\includegraphics[scale=0.44]{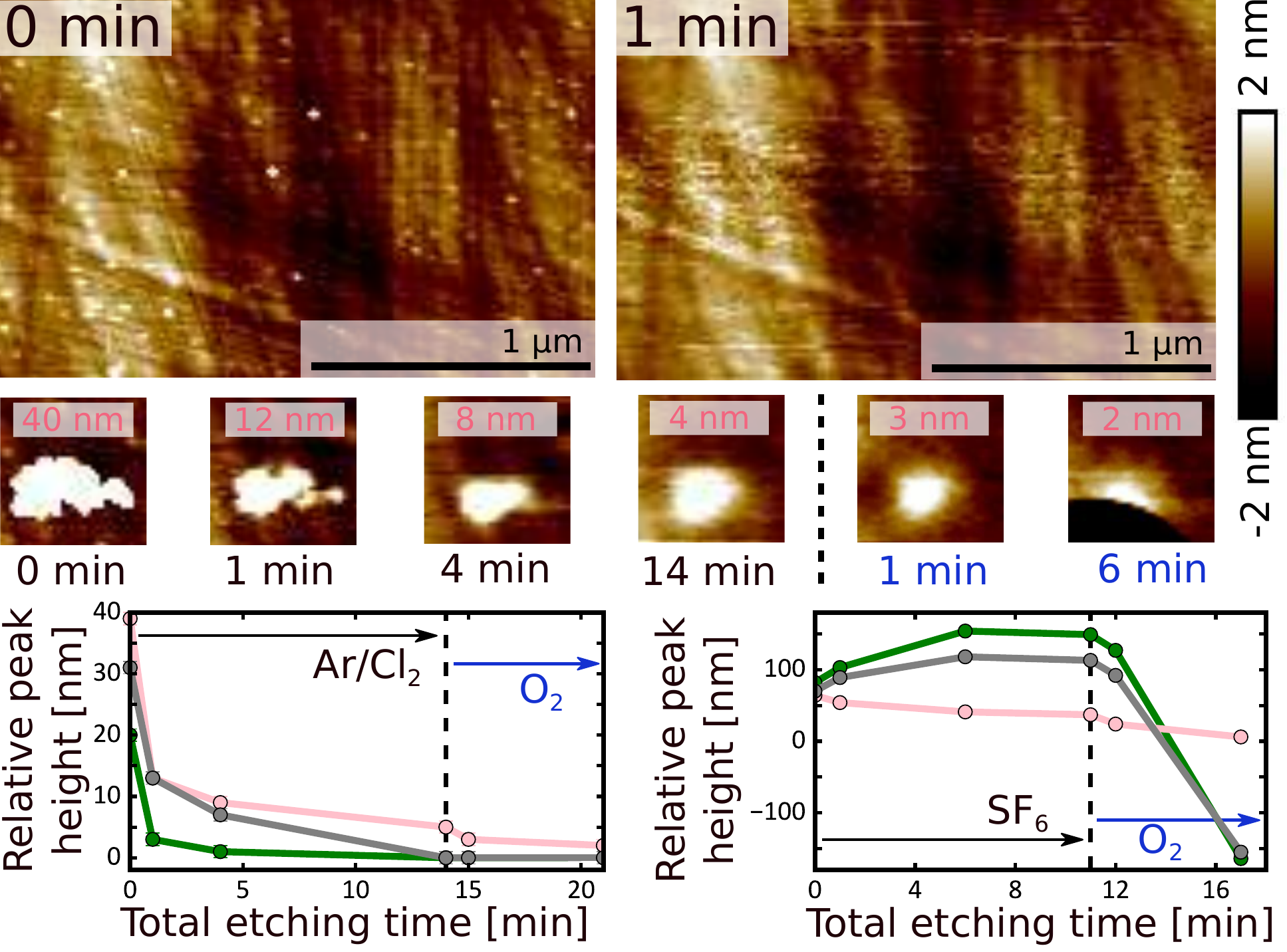}
	\caption{Evolution of particles and surface roughness during etching with Ar/Cl$_2$ of a diamond surface and relative particle height vs.~etching time for SF$_6$ and Ar/Cl$_2$ etching. Top row of AFM images demonstrates that one minute of Ar/Cl$_2$ etching effectively removes small particles. Middle row of AFM images (1 \um by 1 \umdot) shows the evolution of the relative peak height of an initially large particle w.r.t.~the mean surface height as a function of etching time during etching with Ar/Cl$_2$ (black time indication) and subsequent with O$_2$ (blue indication), respectively. Note that the big hole after six minutes of etching with O$_2$ results from etch-induced widening of an initially present hole on the diamond surface, caused by a low-quality diamond growth and polishing process~\cite{supplement}. Bottom graphs compare relative peak height (or hole depth) evolution under ArCl$_2$ (left graph) and SF$_6$ (right graph) pre-etching, followed by O$_2$ etching for different selected particles (see~\cite{supplement} for the full underlying dataset).}
	\label{fgr:arclo2evolution} 
\end{figure}

To reduce the number of particles on the diamond surface to levels lower than after acid cleaning alone~\cite{supplement}, common strategies are to etch under Ar/Cl$_2$~\cite{Lee2008} or SF$_6$~\cite{Challier2018}. However, Ar/Cl$_2$ etching induces Cl contamination on the diamond surface~\cite{Tao2014}, which is suspected to have a detrimental influence on the optical and spin properties of NV centers. Therefore, Ar/Cl$_2$ is often combined with O$_2$ etching~\cite{Maletinsky2012,Chu2014,Riedel2014,Appel2016, Riedel2017a,Challier2018}. Fig.~\ref{fgr:arclo2evolution} confirms that Ar/Cl$_2$ is indeed highly effective in removing particles, and that it can be followed by O$_2$ etching without forming holes on the surface. By comparing the evolution of the relative peak height of particles w.r.t.~the mean of the surrounding diamond surface (see Fig.~\ref{fgr:arclo2evolution}, bottom graphs), we find that Ar/Cl$_2$ is more efficient in removing particles from the diamond surface than SF$_6$~\cite{supplement}: while Ar/Cl$_2$ removes small particles within the first minute of etching and continues to reduce the relative peak height of large debris for longer etching times, it takes longer for small particles to be removed in SF$_6$, and the relative peak height of some larger remaining structures increases during etching in this chemistry, leading to the formation of holes.

\begin{figure}[h!]
	\includegraphics[scale=0.0445]{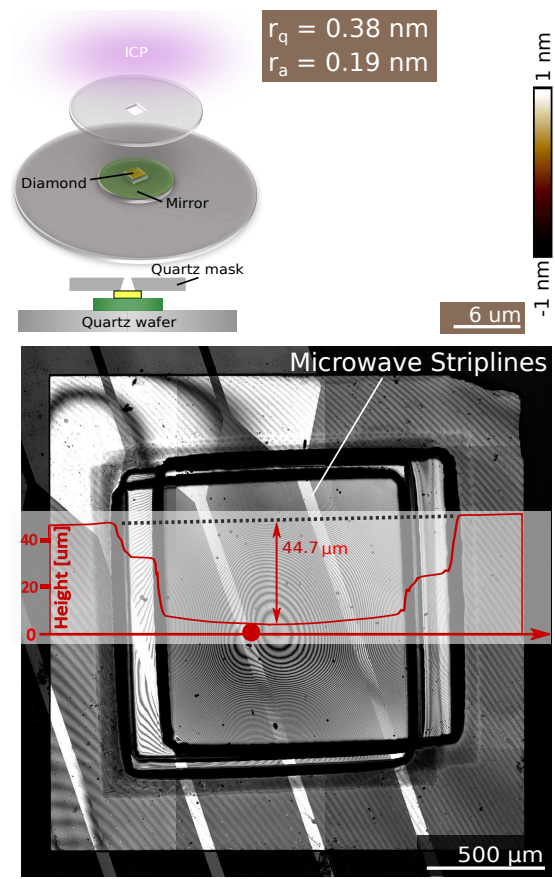}
	\caption{Schematic of the diamond etching setup and resulting membrane profile. (Top left) Setup used for bonded diamond etching. A diamond membrane is bonded to a mirror patterned with gold position markers and striplines, positioned on a fused quartz carrier wafer, and masked from the top with a fused quartz substrate that has a rectangular opening~\cite{Appel2016,supplement}.
		(Bottom) Confocal microscope image of an etched diamond membrane, with three clearly visible offset recesses that result from repeated and shifted partial exposure. The red arrow indicates the profilometer path along which a height profile (red) was taken. It reveals a diamond wedge of 0.14$\degree$ (angle of dotted line w.r.t.~horizontal axis) resulting from the diamond slicing and polishing process, and a maximum etch depth of (44.7 $\pm$ 0.2) \um in the middle region. The red circle corresponds to the region within which most of the data in this paper was taken, with a final thickness of (3.8 $\pm$ 0.2) \umdot. The fringe spacing corresponds to a height change of 84 nm. (Top right) AFM image of the region indicated with a red circle in the middle image, showing a low surface roughness of r$_q$ = 0.38 nm  over an area of 20 \um by 30 \umdot.}
	
	\label{fgr:diamondetchingsetup}
\end{figure}

Using the above etch recipe, we fabricate a sample following the steps in Fig.~\ref{fgr:fabflowandsetup}. For microwave delivery and repeated identification of the measurement area, the sample is bonded via Van der Waals forces to a super-polished mirror patterned with golden striplines and unique position markers before etching~\cite{BogdanovicFab2017}. We have observed that the exposure of mirror material to the plasma leads to severe micromasking on the diamond as etched mirror material is re-deposited on the diamond surface. This results in the formation of holes on the diamond following the same mechanism as discussed above. To mitigate this effect, we use a fused quartz mask for partial exposure of the diamond~\cite{Appel2016} (see Fig.~\ref{fgr:diamondetchingsetup} (Top left)). We thin the sample in three etching steps, each consisting of an Ar/Cl$_2$ pre-etch, followed by a single  O$_2$ step, for a total etch duration of 86 minutes of Ar/Cl$_2$ and 206 minutes of O$_2$ etching. Fig.~\ref{fgr:diamondetchingsetup} (Bottom) shows a confocal microscope image and stylus profilometer height trace of the bonded membrane after the full etching sequence. The geometry of the mask restricts the solid angle of incidence and leads to a position-dependent etch rate. Therefore, the sample height profile shows a curvature in the exposed region, with a thinnest membrane thickness of (3.4 $\pm$ 0.2) \umdot. The red dot indicates the (3.8 $\pm$ 0.2) \um thick area within which the AFM image in Fig.~\ref{fgr:diamondetchingsetup} (Top right) was taken after the full etch. This image confirms a smooth diamond surface ($r_q$ = 0.38 nm over a 20 \um by 30 \um area) even after this prolonged etch sequence.

We characterize the optical properties of NV centers in between etch steps in a confocal microscopy setup $<$ \mbox{10 K} by using a largely automatized measurement sequence to determine their linewidths via photo-luminescence excitation (PLE) scans of the ZPL transitions~\cite{supplement}. After roughly localizing an NV transition in frequency space, we scan a tuneable laser around this frequency, while \mbox{detecting} photons emitted from the NV in the phonon side band (PSB). We repeatedly apply a sequence of a short green laser pulse (to ensure spin and charge state initialization), followed by a red frequency sweep through the expected transition frequency (to map out the specific ZPL transition under dephasing). We do this while constantly applying microwaves to avoid pumping in an optically dark spin state. By performing many scans of this kind, we probe both the effects of spectral diffusion (via a fit to the averaged counts of all scans), as well as the average dephasing width (by fitting each scan individually, and calculating the weighted mean for all fitted linewidths).

\begin{figure} [t!]
	
	\includegraphics[scale=0.34]{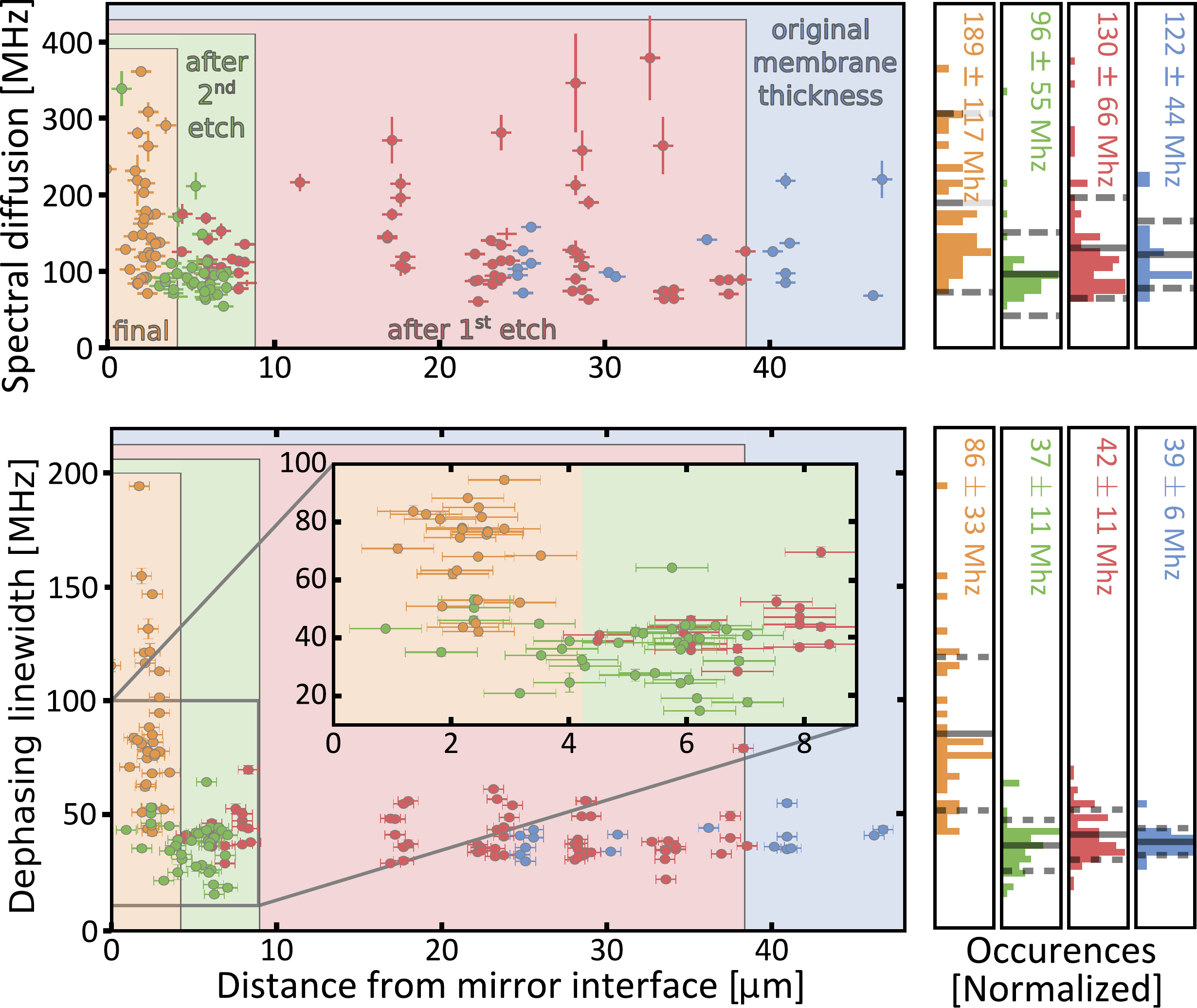}
	\caption{Overview of spectral diffusion (top left panel) and dephasing linewidth (bottom left panel) of NV centers at various distances from the mirror interface for different membrane thicknesses. Right hand side shows normalized histograms of the data on the left, with the black solid (dashed) lines visualizing the mean (standard deviation) of the data for a given membrane thickness, and the colored numbers indicating these values. (Blue) Data before any etching, membrane thickness in measurement region t$_m$ = (47.8 $\pm$ 0.2) \umdot. (Red) Data after first etching of 26 mins Ar/Cl$_2$, followed by 45 mins of O$_2$, t$_m$ = (37.7 $\pm$ 0.2) \umdot. (Green) Data after an additional 30 mins Ar/Cl$_2$ and 138 mins O$_2$ etching, t$_m$ = (10.1 $\pm$ 0.2) \umdot. (Yellow) Data after an additional 30 mins Ar/Cl$_2$ and 23 mins O$_2$ etching, t$_m$ = (3.8 $\pm$ 0.2) \umdot. The horizontal clustering of points is a data acquisition artifact. 
	}
	\label{fgr:opticaloverview}
\end{figure}

Fig.~\ref{fgr:opticaloverview} shows the results of spectral diffusion and dephasing widths as a function of distance from the mirror interface for a total of 155 NV center transitions, stemming from 110 distinguishable NV centers. This data has been acquired at four different steps during membrane thinning and thus membrane thicknesses t$_m$ in the measurement region. While we are not able to track identical NV centers throughout the different etch steps, we make sure to look at the same 20 \um by 20 \um area of diamond for all measurements, apart from the data for the thinnest membrane, for which we additionally included a second region $\sim$ 200 \um from the main measurement area for increased statistics.

The data shows the potential of electron irradiation as a reliable way of introducing coherent NV centers throughout the membrane: for the unetched case, we find an average dephasing width of (39 $\pm$ 6) MHz, and an average spectral diffusion width of (122 $\pm$ 44) MHz. We suspect that laser-power induced linewidth broadening prevents us from resolving lifetime limited values for the dephasing linewidths~\cite{supplement}. Importantly, for membranes etched down to  t$_m$ = (37.7 $\pm$ 0.2) \um and to t$_m$ = (10.1 $\pm$ 0.2) \umdot, we observe that all linewidth averages overlap within statistical uncertainties.

We find roughly two times broader linewidths when characterizing NVs at the final membrane thickness of t$_m$ = (3.8 $\pm$ 0.2) \um in the measurement region; the spectral diffusion width averages to (189 $\pm$ 117) MHz, while the dephasing width averages to (86 $\pm$ 33) MHz. Yet, even for this thickness, we find that 22 out of 37 measured NV centers fulfill our pre-set criteria of dephasing linewidth $<$ 100 MHz and spectral diffusion linewidth $<$ 250 MHz.

The NV ZPL transitions shift with crystal strain: while axial strain results in an overall resonance frequency shift, transverse strain splits the $E_x$ and $E_y$ optical transitions~\cite{Doherty2013}. To determine whether strain in the diamond influences the observed linewidth broadening, we extract the transverse and axial strain from a subset of NV centers via a reconstruction of the NV center Hamiltonian~\cite{supplement}. Although the average axial and transverse strain increase with decreasing NV distance from the mirror interface, the data does not show a dependency of the measured spectral diffusion and dephasing linewidths on strain~\cite{supplement}, suggesting that there is no direct causal relation between the two. Possible causes of increased strain found for NVs after the last etching step include a stressed layer that remains after the membrane polishing process~\cite{Volpe2009,Friel2009,Naamoun2012}, effects due to bonding of the diamond to the mirror (including stress resulting from the difference in thermal expansion coefficients for the diamond and mirror), and lattice-damage induced by reflection of ions from the diamond-mirror interface. Future systematic studies beyond the current work are required to pinpoint the origin unambiguously.

In conclusion, we have demonstrated the fabrication of a diamond sample with a high density of NV centers, introduced through electron irradiation and subsequent high temperature annealing. The surface roughness (r$_q <$ 0.4 nm for a 20 \um by 30 \um area), thickness ($\sim$ 4 \umdot) and NV linewidths ($<$ 100 MHz) of this sample allow for enhanced entanglement generation rates via the Purcell effect in an open, tuneable microcavity setup. Given these sample properties and vibration levels of 0.1 nm rms under pulse-tube operation~\cite{Bogdanovic2017}, we expect an emission of ZPL photons coupled into the fiber mode of 35$\%$~\cite{VanDam2018}, which translates into an entanglement rate enhancement of two orders of magnitude. This would allow one to form a quantum repeater beating direct transmission~\cite{Briegel1998,Rozpedek2018,Rozpedek2018a}, signaling the surpassing of a fundamental milestone on the route to building a quantum network. 

\bibliography{main_arxiv}

\section*{Acknowledgements}

The authors thank Marinus Hom, Wybe Roodhuyzen and Ferdinand Grozema for electron irradiation of diamonds, Kevin Chang, Michael Burek, Daniel Riedel and Erika Janitz for helpfull nanofabrication discussions, Charles de Boer, Eugene Straver, Marc Zuiddam and Jasper Flipse for fabrication assistance, Airat Galiullin for developing early versions of the measurement scripts and Matthew Weaver and David Hunger for careful reading of our manuscript. This research was supported by the Early Research Programme of the Netherlands Organisation for Applied Scientific Research (TNO), and by the Top Sector High Tech Systems and Materials. We furthermore acknowledge financial support from the Netherlands Organisation for Scientific Research (NWO) through a VICI grant, and from the European Research Council through an ERC Consolidator Grant.

\section*{Supporting information}
Details on sample preparation and etch recipe development. Fabrication of sample described in main text. NV linewidth characterization methods. Laser-power induced linewidth broadening. NV depth conversion factor and error analysis. Correlations of NV center strain and linewidths.

\section*{Competing Financial Interests}
The authors declare no competing financial interests.

\end{document}